\documentclass[letterpaper, 10 pt, conference]{ieeeconf}

\newcommand{\R}{\mathbb{R}}

\usepackage{float}
\usepackage{amssymb}
\usepackage{amsmath}

\usepackage{bbold}

\usepackage{url}          
\usepackage{hyperref}

\usepackage{caption}
\usepackage{comment}
\usepackage[ruled]{algorithm2e}
\usepackage{float}
\usepackage{multicol}
\usepackage{color}
\usepackage{graphicx}
\newcounter{multifig}
\usepackage{multirow}

\usepackage{cuted} 

\usepackage{lipsum}

\usepackage{mathtools}

\usepackage[caption=false, font=footnotesize]{subfig}

\newtheorem{theorem}{Theorem}[section]

\newtheorem{observation}[theorem]{Observation}

\newtheorem{remark}[theorem]{Remark}

\newtheorem{assumption}[theorem]{Assumption}
\newtheorem{problem}[theorem]{Problem}

\title{\LARGE \bf Derandomized Distributed Multi-resource Allocation with Little Communication Overhead}

\author{Syed Eqbal Alam \\ CIISE,
		Concordia University, \\
		Montreal, Quebec, Canada,\\email:  sy\_al@encs.concordia.ca
		\\
	\and Robert Shorten\\ School of Electrical, Electronic
		and Comm. Engineering, \\University College Dublin, Dublin, Ireland,\\email: robert.shorten@ucd.ie
		\\
	\and Fabian Wirth \\ Faculty of Computer Science and Mathematics,\\ University
		of Passau, Passau, Germany 
		\\
	\and Jia Yuan Yu \\ CIISE,
	Concordia University, \\
	Montreal, Quebec, Canada,\\ email: jiayuan.yu@concordia.ca}

\begin{document}
	
	\maketitle
	\thispagestyle{empty}
	\pagestyle{empty}
	\begin{abstract}
	We study a class of distributed optimization problems for multiple shared resource allocation in Internet-connected devices. We propose a derandomized version of an existing stochastic additive-increase and multiplicative-decrease (AIMD) algorithm. The proposed solution uses one bit feedback signal for each resource between the system and the Internet-connected devices and does not require inter-device communication. Additionally, the Internet-connected devices do not compromise their privacy and the solution does not dependent on the number of participating devices. In the system, each Internet-connected device has private cost functions which are strictly convex, twice continuously differentiable and increasing. We show empirically that the long-term average allocations of multiple shared resources converge to optimal allocations and the system achieves minimum social cost. Furthermore, we show that the proposed derandomized AIMD algorithm converges faster than the stochastic AIMD algorithm and both the approaches provide approximately same solutions.
	\end{abstract}
	
	\let\thefootnote\relax\footnotetext{This preprint is to appear in Allerton Conference on Communication, Control, and Computing, Urbana-Champaign, IL, USA, October, 2018. \\ The work is partly supported by Natural Sciences and Engineering Research Council of Canada grant no. RGPIN-2018-05096.}	
	\section{Introduction}
\begin{figure}
		\centering
		\includegraphics[width=0.6\textwidth,clip=true,trim=8.5cm 7.05cm .05cm 4.8cm]{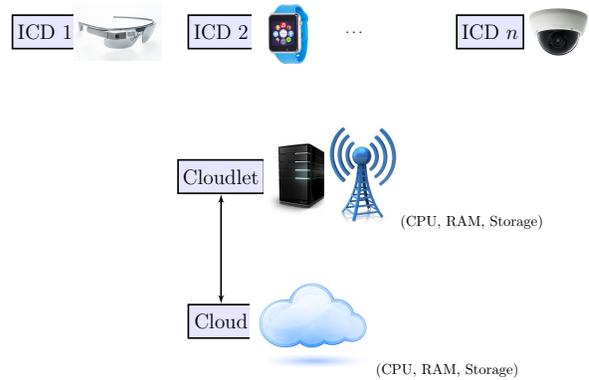}
	
		\caption{A three-tier architecture of ICDs --- Cloudlets --- Cloud: Internet-connected devices offload their tasks on Cloudlets. They receive computing resources such as CPU, memory, storage, etc., from Cloudlets with little latency. Larger and latency tolerant tasks can be offloaded on Cloud.}
		\label{Sys_diag}
	\end{figure}

	The number of Internet-connected devices, for example, smart phones, smart watches, fitness trackers, connected cars, cameras, etc., is increasing very rapidly \cite{Columbus2017}. Such devices are constrained by computational resources and battery life \cite{Atzori2010,Fuqaha2015}, by providing additional shared resources for task offloading, these devices can be used to build many emerging smart applications.  Some of the representative smart applications are, Apple Siri, Google assistant, IBM Watson, Amazon Alexa, etc., they use Cloud technologies for task offloading. Because Clouds are usually hosted at faraway locations worldwide from the Internet-connected devices (ICDs), offloading the tasks on the Clouds incur delay, which is not suitable for many latency-sensitive mobile applications, for example, cognitive assistance \cite{Ha2014}. Interested readers can find some interesting cognitive assistance applications in \cite{Chen2017_1}. Such applications collect the data through sensors of wearable devices, offload them on mini Cloud data-centers called {\em Cloudlets} \cite{Satyanarayanan2009} for processing and receive the processed information in real time, which assists the cognitive impaired people. Basically, Cloudlets are the enabling technology for computing resource intensive and latency-sensitive mobile applications \cite{Satyanarayanan2017}. It stays in close proximity to the Internet-connected devices usually at one hop \cite{Bruin2018}. A recent survey of such technologies can be found in \cite{Shi2016}. The Internet-connected devices receive the resources on-demand from a Cloudlet to perform interactive tasks. However, the latency tolerant tasks can be offloaded on the Cloud. Diagram of such a system is presented in Figure \ref{Sys_diag}. A similar three-tier architecture is presented in \cite{Song2017}, where a human-centric real-time positioning system using wearable devices is proposed. The resource provisioning in a Cloudlet is dynamic and the Internet-connected devices receive resources as a virtual machine (VM). Interested readers can find details of on-demand VM provisioning in \cite{Echeverria2014}. Since, each device works for different purpose, it requires different amount of shared resources. We assume that the VMs are customized, which are created based on the resource requirement of ICDs.

	  Due to the increase in number of Internet-connected devices, optimal allocation of shared resources with aim to achieve social optimum with little communication overhead is a challenging problem. It becomes more challenging to achieve social optimum when the ICDs do not communicate their information to other ICDs. As a step in this direction, Alam et al. \cite{Eqbal2017} propose a distributed stochastic additive increase and multiplicative decrease (AIMD) algorithm to allocate multiple shared resources to ICDs. The algorithm is a distributed stochastic algorithm and involves very little communication overhead. Based on this work, we propose a distributed and deterministic AIMD algorithm, which is a derandomized version of it. The proposed algorithm has very little communication overhead and converges much faster than the stochastic AIMD algorithm \cite{Eqbal2017}. Interested readers can find recent works on resource allocation in \cite{Angelakis2016, Lu2016}.
	In the proposed algorithm, we assume that each ICD has its own cost function and it does not share its cost function or resource allocation history with other participating ICDs. 
	
	Additive increase and multiplicative decrease (AIMD) algorithm is used in transmission control protocol (TCP) for congestion control \cite{Chiu1989, Corless2016}, Wirth et al. \cite{Wirth2014} introduced the probabilistic intent in it to solve a class of optimization problems. It is further generalized by Alam et al. \cite{Eqbal2017} to solve multi-variate optimization problems. As a background, in stochastic AIMD algorithm \cite{Eqbal2017}, the devices keep increasing their demand of each resource linearly by a constant called {\em additive increase factor}, until they receive a {\em capacity event} notification from the control unit. The control unit sends a capacity event notification for each resource to all ICDs to notify that the total demand of the resource has exceeded its capacity. After receiving this notification, the devices respond in a probabilistic way to decrease their resource demands abruptly by a constant called {\em multiplicative decrease factor}. The devices again start increasing their demand linearly until they receive the next capacity event notification, this process repeats over time and repeats for all the resources in the system. We modify this stochastic AIMD algorithm and propose a deterministic algorithm to allocate multiple shared resources. We assume that there is a Cloudlet which hosts several computing resources and ICDs offload their tasks on it. The aim of the system is, to allocate resources hosted on Cloudlet optimally and achieve social optimum cost. The proposed deterministic AIMD algorithm, works as follows, at the start of the algorithm the control unit of the Cloudlet broadcasts a set of parameters to all the ICDs in the system, each ICD has its own cost function and it does not share it with other ICDs. After joining the system, ICDs keep increasing the resource demand linearly by a positive constant called {\em additive increase factor} until they receive the {\em capacity event} notification from the control unit of the Cloudlet. After receiving capacity event notification from the control unit, they abruptly decrease their demand in a deterministic way, based on multiplicative decrease factor, average resource allocation and derivative of the cost functions. The devices can start increasing their demand again until they receive the next capacity event notification from the control unit. This process repeats over time and repeats for every resource in the system. 
	By doing so, the long-term average allocations converge to optimal allocations for each resource and the system achieves a social optimum value. We observe empirically that the proposed deterministic AIMD algorithm converges faster than stochastic AIMD algorithm. Furthermore, we observe that, both the approaches provide approximately the same results over time. The long-term average allocations are approximately equal to the optimal allocation values obtained by solving the same optimization problem in a centralized setting.

	The following are three main contributions in this paper.
	First, we propose a distributed, private and deterministic resource allocation algorithm to allocate multiple shared resources to ICDs. The algorithm is a derandomized version of the stochastic AIMD algorithm \cite{Eqbal2017}. 
	Second, we describe simulation settings in detail and to verify the efficacy of our results, we compare the results with stochastic AIMD algorithm. We also compare the results of deterministic AIMD with the results obtained by solving the same optimization problem in a centralized setting.
	Third, we present a use case of a tourist attraction center, where we assume heterogeneous Internet-connected devices, such as a set of wearable devices, surveillance cameras, etc. These Internet-connected devices receive on-demand shared resources from the Cloudlets for task offloading. Such facilities can be useful in realizing the potential of Internet-connected devices in emerging applications such as, assisting the tourists with physical disability, uploading the pictures of scenes on social media, getting real time detailed information about a painting in a museum, real time foreign language translation, etc. 
		
	The rest of the paper is organized as follows, in Section \ref{prob_form} we describe the notations used and formulate the optimization problem. In Section \ref{algo_stoc}, we present the details of stochastic AIMD algorithm. In Section \ref{algo_det} we describe and formally present the proposed distributed deterministic algorithm for multi-resource allocation. In Section \ref{exp_results}, we present a use case of tourist attraction center in which heterogeneous Internet-connected devices are assumed. In this section, we also describe the simulation settings and numerical results. Section \ref{conc} describes the open problems and future applications.  
	\section{Problem formulation} \label{prob_form}
	Suppose that a Cloudlet hosts a pool of shared resources, e.g., CPUs, GPUs, RAM, storage, network bandwidth, etc. For the sake of generality we assume that there  are $m$ shared resources $R^1, R^2, \ldots, R^m$ with capacity $C^1, C^2, \ldots, C^m$,
	respectively. Let there are $n$ heterogeneous Internet-connected devices, $ICD_1$, $ICD_2$, $\ldots, ICD_n$, such as smart wearable devices, cameras, etc. The Cloudlet has a control unit, which coordinates with the Cloudlet and the participating Internet-connected devices (ICDs). We index ICDs with $i \in \{1,2,\ldots,n\}$ and use $j \in \{1, 2, \ldots, m\}$ to index the resources hosted on the Cloudlet. Let $\mathbb{N}$ denotes the set of natural numbers. We use $k$ to index time steps, where $k \in \mathbb{N}$. Furthermore, we assume that each ICD has a cost function $f_i: \mathbb{R}^m \to \mathbb R$ which associates a cost to a certain allotment of resources. An ICD demands the amount of shared computing resources from the Cloudlet according to its need and the application it performs. The aim of the system is that each ICD receives the optimal allocation and the system achieves a minimum social cost. Now, we list the following assumptions about the cost functions of each ICD.
	\begin{assumption}
		We assume that the cost function $f_i$ is twice continuously differentiable, convex, and
		increasing. We also assume that a device does not share its resource requirement or cost function with other devices.
	\end{assumption}
	Now, for all $i$ and $j$, we denote by $x_i^j \in \mathbb{R}_+$ the
	amount of resource $R^j$ allotted to ICD $i$. 
	We aim to solve the following optimization problem:
	\begin{problem} \label{obj_fn1}
		\begin{align*} 
		\begin{split}
		\min_{{x}^1_1, \ldots, {x}^m_n} \quad &\sum_{i=1}^{n} f_i(x^1_i, x^2_i,
		\ldots, x^m_i),    		\\
		\mbox{subject to} \quad
		&\sum_{i=1}^{n} x^j_i = C^j, \quad j \in \{1,2,\ldots,m\},		\\
		&x^j_i \geq 0, \quad i \in \{1,2,\ldots,n\}, \ j \in \{1,2,\ldots,m\}.
		\end{split}
		\end{align*}
	\end{problem}
	We denote the solution to the minimization problem by
	$x^{*} \in \mathbb{R}_+^{nm}$, where $x^* = [x_1^{*1}, \ldots, x_n^{*m}]^T$, where $T$ denotes the transpose. As described in \cite{Wirth2014, Syed2018}, because of the compactness of the constraint sets and strict convexity of the cost functions, we say that there exists a unique optimal solution of Problem \ref{obj_fn1}.
	Now, let $x_i^j(k)$ and $\overline{x}^j_i(k)$ denote the amount of instantaneous allocation and the amount of average allocation of resource $R^j$ of ICD $i$ at the time step $k$, respectively. For all $i$, $j$ and $k$, the average allocation is calculated as follows,
	\begin{align} \label{average_eqn}
	\overline{x}^j_i(k)=\frac{1}{k+1} \sum_{\ell=0}^k x^j_i(\ell).
	\end{align}
	We assume here that an ICD can obtain any amount of resource in $[0,C^j]$. Our aim here is to propose a distributed, private and deterministic iterative scheme to allocate multiple shared resources to ICDs, such that the long-term average allocations of resources converge to the optimal allocations. Let $\overline{x}(k) \in \mathbb{R}_+^{nm}$, where $\overline{x}(k) = [\overline{x}_1^{1}(k), \ldots, \overline{x}_n^{m}(k)]^T$. Then, we aim to achieve,
	\begin{align} \label{eq:longtermopt}
	\overline{x}(k) \to x^{*}, \text{ when } k\to \infty.
	\end{align}
	Let $\nabla_j f_i(.)$ denotes the (partial) derivative of $f_i(.)$ with respect to resource $R^j$, for all $j$. 
	Based on the analysis of  \cite{Syed2018}, we say that the derivatives of cost functions of all ICDs for a particular resource should make a consensus to achieve optimality for Problem \ref{obj_fn1}.
	\section{Stochastic AIMD Algorithm} \label{algo_stoc}
	In this section, we briefly describe the stochastic AIMD algorithm \cite{Eqbal2017}, which is a distributed, iterative and randomized multi-resource 
	allocation algorithm based on AIMD algorithm. The stochastic AIMD algorithm consists of two phases: {\em additive increase} and {\em multiplicative decrease}. In additive increase phase, the ICDs or agents keep increasing their demands of a resource linearly by a constant called {\em additive increase factor} $\alpha^j \in \R_+$ until they receive a {\em capacity event} notification from the control unit. The control unit broadcasts the capacity event to all ICDs when their demands exceed the capacity of the resource, i.e., $\sum_{i=1}^{n} x_i^j(k) > C^j$. We write the additive increase phase as follows,
	\begin{align*}
	x_i^j (k+1) = x_i^j (k) + \alpha^j.
	\end{align*}
	After receiving this notification, the ICDs respond in a probabilistic way to decrease their resource demand by a constant called {\em multiplicative decrease factor} $\beta^j \in [0,1)$, this phase is called {\em multiplicative decrease} phase. Let the probability of ICD $i$ at time step $k$ for resource $R^j$ is denoted by $\lambda_i^j(k)$. Let $\mathbb{1}_i^j(k)$ be an indicator function, defined as follows,
	\begin{align*}
	\mathbb{1}_i^j(k) =
	\begin{cases}
	1 & \text{w.p. } \lambda_i^j(k)\\
	0 & \text{w.p. } 1- \lambda_i^j(k).
	\end{cases}
	\end{align*}
	Then, we write the multiplicative decrease phase as follows,
	\begin{align*}
	x_i^j(k+1) = \mathbb{1}_i^j(k) \beta^j x_i^j(k) + \big(1-\mathbb{1}_i^j(k)\big) x_i^j(k).
	\end{align*}
	
	ICDs calculate probability $\lambda_i^j(k)$ using their average allocations, derivative of their cost functions and normalization factor received from the control unit. After multiplicative decrease phase, the ICDs again start increasing their demand linearly by $\alpha^j$ until they receive the next capacity event notification, this process repeats over time and repeats for all the resources. By doing so, the long-term average allocations converge empirically to optimal allocations and hence the solution provides a social minimum cost.

\section{Deterministic AIMD Algorithm} \label{algo_det}
\begin{figure}
		\includegraphics[width=0.54\textwidth,clip=true,trim=7.9cm 9cm 1cm 4.8cm]{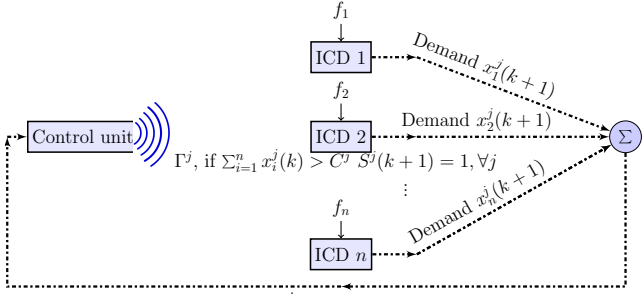}
	
		\caption{Block diagram of the multi-resource allocation deterministic AIMD model.}
		\label{Diag_DAIMD}
	\end{figure}

	The proposed algorithm is a distributed, private, iterative and deterministic AIMD algorithm. It is used to allocate multiple shared resources to Internet-connected devices. The algorithm is deterministic in the sense that it does not involve randomness in resource allocation phases. The block diagram of the system is presented in Figure~\ref{Diag_DAIMD}.  We assume that there is a Cloudlet which hosts computing resources and the control unit is a sub-system of the Cloudlet. Let $\alpha^j \in \R_+$ be the {\em additive increase factor}, $\beta^j \in [0,1)$ be the {\em multiplicative decrease factor} and $\Gamma^j \in \R_+$ be the normalization factor of resource $R^j$, for all $j$. Let $S^j(k) \in \{0,1\}$ denotes \emph{capacity event} at time step $k$ for resource $R^j$,
	for all $j$ and $k$. At the start of the system, control unit initializes the parameters $\Gamma^j$, $\alpha^j$ and $\beta^j$ with desired values. It also initializes the capacity event $S^j(0)$ with $0$. After initialization, the control unit broadcasts $\Gamma^j$, $\alpha^j$, $\beta^j$ and $S^j(0)$ to all the participating ICDs, for all resources.
	 Each ICD runs its own algorithm to demand shared resources. 
	 
	 The algorithm of an ICD works as follows, after joining the system, an ICD starts increasing its demand of shared resources linearly until it receives a capacity event notification from the control unit. The control unit broadcasts this notification when the total demand of a resource exceeds the capacity of the resource. After receiving this capacity event, the ICD decreases its demands abruptly and in a deterministic way. It does so by using multiplicative decrease factor $\beta^j$ and its {\em scaling factor} $\lambda^j_{i}(k) \in (0,1)$. The control unit updates and broadcasts the capacity event $S^j(k+1)=1$ when the total demand $\sum_{i=1}^n x_i^j(k)$ of resource $R^j$ exceeds the capacity $C^j$ at time step $k$, i.e.,
	\begin{align*} 
	S^j(k+1) =
	\begin{cases}
	0 \quad \text{if } \sum_{i=1}^{n} x_i^j(k) \leq C^j \\
	1 \quad \text{if } \sum_{i=1}^{n} x_i^j(k) > C^j.
	\end{cases}
	\end{align*} 
	Now, we describe the deterministic AIMD algorithm formally, it has following two phases, {\em additive increase (AI)} and {\em multiplicative decrease (MD)}.
	\begin{itemize}
		\item[(i)] {\em Additive increase (AI)}: In this phase, an ICD keeps increasing its resource demand linearly by additive increase factor $\alpha^j$ until it receives a  capacity event signal $S^j(k) = 1$ from the control unit of the system. The control unit broadcasts the capacity event to notify the ICDs that their demand has exceeded the capacity of the resource. We describe the AI phase as follows, 
		\begin{align*} 
		x_i^j (k+1) = x_i^j (k) + \alpha^j.
		\end{align*}
		\item[(ii)] {\em Multiplicative decrease (MD)}: In this phase, after receiving the capacity event signal from the control unit, the ICDs reduce their demands synchronously in a deterministic way. To do so, they use the multiplicative decrease factor $\beta^j \in [0,1)$ and scaling factor  $\lambda^j_{i}(k) \in (0,1) $. The value of scaling factor $\lambda^j_{i}(k)$ is calculated according to \eqref{prob_x}. The multiplicative decrease phase is described as follows,
		\begin{align*} 
		x_i^j (k+1) = \Big (\lambda^j_{i}(k) \beta^j  + \big(1 -\lambda^j_{i}(k) \big) \Big)x^j_i(k). 
		\end{align*} 
	\end{itemize}
	After abruptly reducing the resource demand, the ICDs again start to increase their demand linearly until they receive the next capacity event, this process repeats over time. 
	 
	 Now, let $\Gamma^j$ be the normalization constant received by an ICD $i$ for resource $R^j$ from the control unit. It is used to keep scaling factor $\lambda^j_{i}(k) \in (0,1)$. The scaling factor $\lambda^j_{i}(k)$ of ICD $i$ depends on its average resource allocation, the derivative of its cost function with respect to resource $R^j$ and normalization constant $\Gamma^j$. For all $i$, $j$ and $k$, we calculate scaling factor as follows,
	 \begin{align} \label{prob_x} \lambda^j_{i}(k) = \Gamma^j  
	 \frac{{\nabla_j} f_i(\overline{x}^1_i(k), \ldots,
	 	\overline{x}^m_i(k))}{\overline{x}^j_i(k)}.
	 \end{align}
	 Now, let $\mathcal{F}$ be the set of convex, twice continuously differentiable and increasing functions. Then, the control unit calculates $\Gamma^j$ as follows, 
	 \begin{align*} 
	 \Gamma^j = 
	 \inf_{x^1,\ldots,x^m \in \mathbb{R}_+,  f \in
	 	\mathcal{F}}
	 \Big(\frac{x^j}{\nabla_j f(x^1,
	 	x^2, \ldots, x^m)} \Big), \text{ for all } j.  
	 \end{align*}
Notice that the method of calculating scaling factor $\lambda^j_{i}(k)$ and normalization factor $\Gamma^j$ is same as described in \cite{Eqbal2017}. For clarity,  we use the term scaling factor for $\lambda^j_{i}(k)$ instead of probability.
	\begin{algorithm}  \SetAlgoNoLine Input:
		$C^{j}$, for $j \in \{1,2,\ldots,m\}$.
		
		Output:
		$S^{j}(k)$, for $j \in \{1,2,\ldots,m\}$, $k \in
		\mathbb{N}$.
		
		Initialization: $S^{j}(0) \leftarrow 0$, for $j \in \{1,2,\ldots,m\}$,
		\begin{align*}
		\Gamma^j \leftarrow
		\inf_{x^1,\ldots,x^m \in \mathbb{R}_+,  f \in
			\mathcal{F}}
		\Big(\frac{x^j}{\nabla_j f(x^1,
			x^2, \ldots, x^m)} \Big),
		\end{align*}
			broadcast $\Gamma^{j}$, for $j \in \{1,2,\ldots,m\}$;

		\ForEach{$k \in \mathbb{N}$}{
			
			\ForEach{$j \in \{1,2,\ldots,m\}$}{
				\uIf{ $\sum_{i=1}^{n} {x}_i^{j}(k) > C^{j}$}{
					$S^{j}(k+1) \leftarrow 1
					$\;
					
					broadcast $S^{j}(k+1)$;	
				}
				
				\Else{$S^{j}(k+1) \leftarrow 0$\;}
			}
		}
		\caption{Algorithm of control unit}
		\label{algoCUDet}
	\end{algorithm}
	The algorithm of the control unit is presented in Algorithm \ref{algoCUDet} and the distributed deterministic algorithm for each ICD is described in Algorithm \ref{algoDem12}. 
	\begin{algorithm}  \SetAlgoNoLine Input:
		$S^{j}(k)$,  $k \in \mathbb{N}$ and $\Gamma^j$, for $j \in \{1,2,\ldots,m\}$.
		
		Output:
		$x^j_i(k+1)$, for $j \in \{1,2,\ldots,m\}$, $k \in
		\mathbb{N}$.
		
		Initialization: $x^j_i(0) \leftarrow 0$ and
		$\overline{x}^j_i(0) \leftarrow x^j_i(0)$, for
		$j \in \{1,2,\ldots,m\}$;
		
		\While{ICD $i$ participates at $k \in \mathbb{N}$}{
			
			\ForEach{$j \in \{1,2,\ldots,m\}$}{

				\uIf{$S^{j}(k) = 1$}{
					$ \lambda^j_{i}(k) \leftarrow \Gamma^j
					\frac{{\nabla_j} f_i \left ( \overline{x}^1_i(k),
						\overline{x}^2_i(k), \ldots, \overline{x}^m_i(k)
						\right)}{\overline{x}^j_i(k)}$;
								\newline
								\newline
					$x^j_i(k+1) \leftarrow \Big (\lambda^j_{i}(k) \beta^j   +\big(1-\lambda^j_{i}(k)\big) \Big)x^j_i(k) $;
					
				} \Else{
					$x^j_i(k+1) \leftarrow x^j_i(k) + \alpha^j$; }
				
				$\overline{x}^j_i(k+1) \leftarrow \frac{k+1}{k+2}
				\overline{x}^j_i(k) + \frac{1}{k+2} x^j_i(k+1);$
		} }
		
		\caption{Deterministic AIMD algorithm of ICD $i$ (D-AIMD $i$)}
		\label{algoDem12}
	\end{algorithm}
	By following the approach with appropriate values of $\lambda^j_{i}(k)$, $\alpha^j$, $\beta^j$ and $\Gamma^j$, the long-term average resource allocations converge to optimal allocations, i.e., $\overline{x}_i(k) \to x_i^{*}$, when $k \to \infty$ and hence the system achieves a minimum social cost.
	
	Notice that, in additive increase (AI) phase, stochastic as well as deterministic AIMD algorithms follow same allocation rule. However, in multiplicative decrease (MD) phase, they follow different allocation rules. The MD phase in stochastic AIMD algorithm is probabilistic, an ICD responds the capacity event asynchronously --- either it reduces its demand multiplicatively by $\beta^j$ or it does not respond the capacity event, whereas in deterministic AIMD, MD phase is deterministic and all the participating ICDs back-off synchronously using $\beta^j$ and $  \lambda^j_{i}(k)$.
	
	\begin{observation}
		Based on the empirical results we observe that the average allocations converge to optimal allocations. We also observe that the deterministic AIMD algorithm converges much faster than the stochastic AIMD algorithm.
	\end{observation} 
	\begin{remark} [Communication overhead] Because of the faster convergence of deterministic AIMD algorithm the number of capacity events (in bits) required to reach consensus of derivatives is less than that of stochastic AIMD algorithm. As described in \cite{Eqbal2017} for $m$ resources in the system, the communication overhead is $\sum_{\ell=0}^{k} \sum_{j=1}^{m} S^{j}(\ell)$ bits until $k^{th}$ time step, where $S^{j}(\ell) \in \{0,1\}$. In the worst case scenario this will be $mk$ bits until time step $k$. Additionally, in both the algorithms, the communication complexity is independent of the number of participating ICDs. Furthermore, failure of an ICD does not affect the performance of the system.
	\end{remark}
	%
	%
	
	\section{Experiments} \label{exp_results}
	In this section, we present the simulation settings and results of the deterministic  AIMD algorithm and compare it with the results of stochastic AIMD algorithm, which  is simulated using same parameters. We observe that in both approaches, the average allocations of a resource converge at approximately same value for each ICD. We also observe that deterministic AIMD converges faster than the stochastic AIMD. 
	
	Now, suppose that there are $60$ Internet-connected devices such as Google glasses, smart watches and cameras being used in and around a place of tourist attraction. Let us assume that the management of the place has installed a Cloudlet there and aims to allocate resources optimally to all participating ICDs and achieve a social minimum cost. We assume further that the Cloudlet hosts three shared computing resources, RAM $R^1$, CPU cycles $R^2$ and disk storage $R^3$. Let the capacities of these resources be $C^1 = 32$ GB, $C^2 = 20$ GHz and $C^3 = 250$ GB, respectively. For the convenience of scalability of parameters, we suppose that $10$ GB of storage is represented by $\mathrm{GB^D}$, then we write $C^3 = 25$ $\mathrm{GB^D}$. The Cloudlet has a control unit that notifies the ICDs when the demands exceed the capacity of a resource. At the start of the system, the control unit broadcasts the parameters $\alpha^1 = 0.025$ GB, $\alpha^2 = 0.02$ GHz, $\alpha^3 = 0.0225$ $\mathrm{GB^D}$, $\beta^1 = 0.7$, $\beta^2 = 0.85$ and $\beta^3 = 0.75$ to all the participating ICDs. It also broadcasts the normalization factors $\Gamma^1 = \Gamma^2 = \Gamma^3 = 1/90$ for resources $R^1, R^2$ and $R^3$, respectively. 
	
	Let $a_i, b_i,$ and $c_i$ represent the cost of RAM, CPU cycles and disk storage, respectively and $d_i$ represents any other costs incurred. Let $X = \{1, 2, \ldots, 25\}$, $Y= \{1, 2, \ldots, 20\}$, $Z = \{1, 2, \ldots, 15\}$ and $U = \{1, 2, \ldots, 10\}$ be uniformly distributed random variables. We use $a_i \in X, b_i \in Y, c_i \in Z$ and $d_i \in U$ as the numerical values of these uniformly distributed random variables in the simulation. Let for all $i$, the cost functions of ICD $i$ is calculated as follows,	
		\begin{strip}
		\begin{align} \label{cost_fn}
		f_{i}(x_i^1, x_i^2, x_i^3) =
		\begin{cases}
		a_i\big((x_i^1)^2 + \frac{1}{2}(x_i^1)^4 \big) + b_i \big(2(x_i^2)^4 + \frac{1}{2}(x_i^2)^6 \big) + c_i \big( (x_i^3)^2 + \frac{1}{4} (x_i^3)^4 \big) + \frac{1}{8}d_i(x_i^3)^8  & \text{ w.p. } 1/3 \\
		a_i(x_i^1)^2 + b_i \big ( (x_i^2)^2 + \frac{1}{2}(x_i^2)^4 \big ) + \frac{3}{2} c_i(x_i^3)^4 & \text{ w.p. } 1/3 \\
		\frac{1}{3} a_i(x_i^1)^6  + b_i(x_i^2)^2 + c_i(x_i^3)^2 + d_i \big(\frac{1}{6}(x_i^2)^6 + \frac{1}{8}(x_i^3)^4 \big) & \text{ w.p. } 1/3. 
		\end{cases}
		\end{align}
	\end{strip}
	\begin{figure*}[h] 
		\centering
		\subfloat[]{%
			\includegraphics[width=0.33\linewidth]{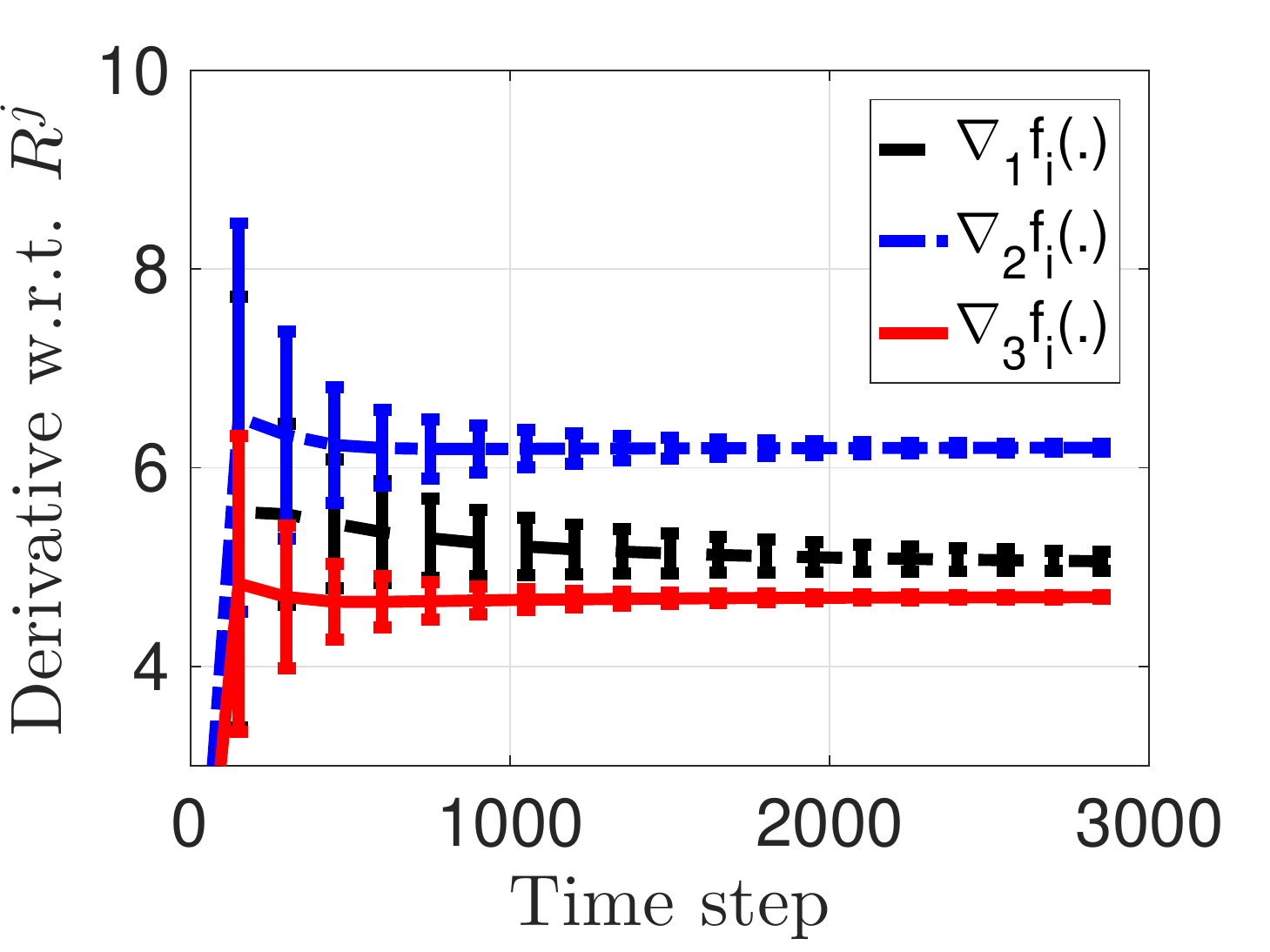}}
		\label{err_grad_rand_alpha2}\hfill
		\subfloat[]{%
			\includegraphics[width=0.33\linewidth]{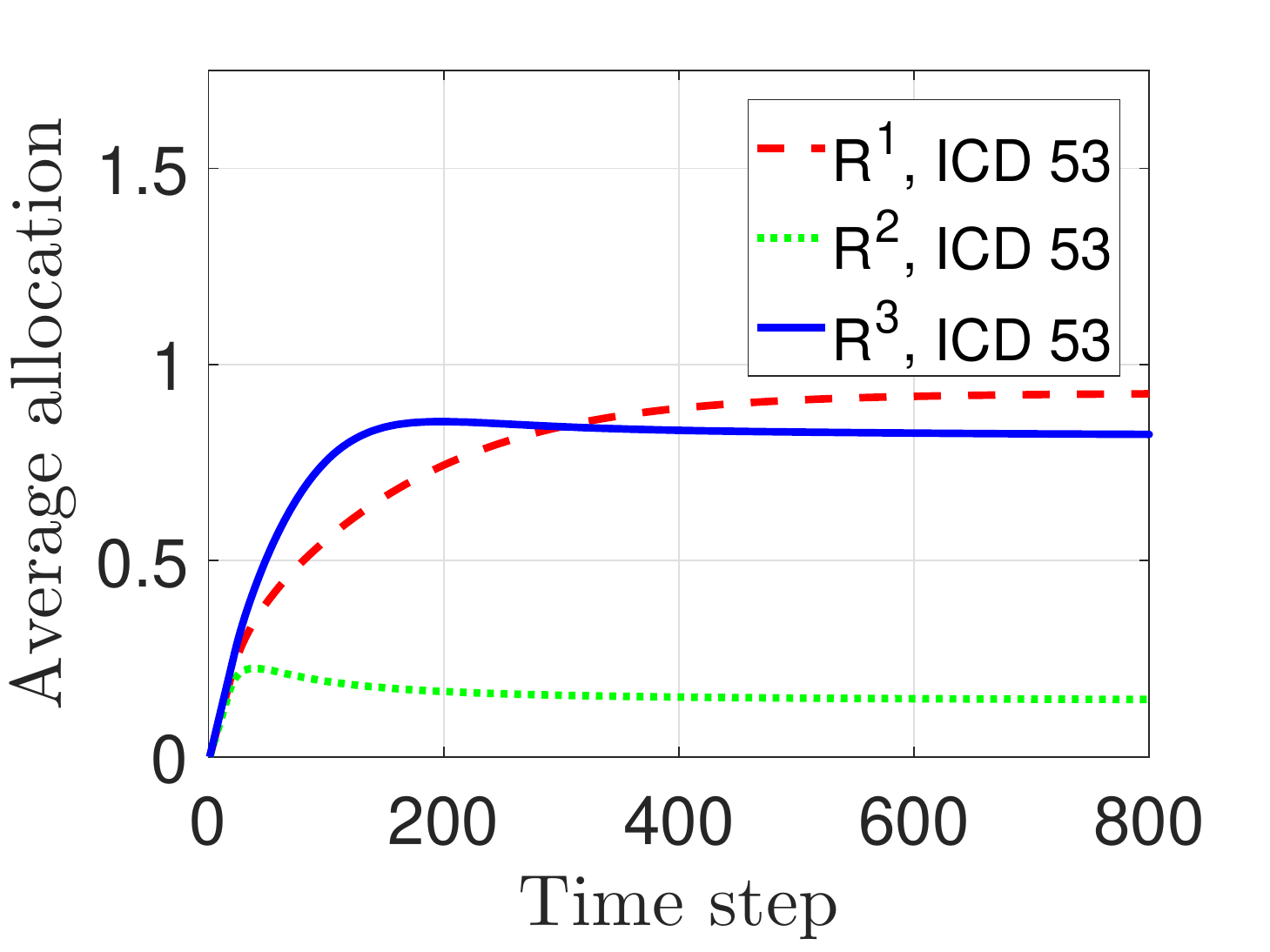}}
		\label{avg_AIMD_rand_alpha2}\hfill
		\subfloat[]{%
			\includegraphics[width=0.33\linewidth]{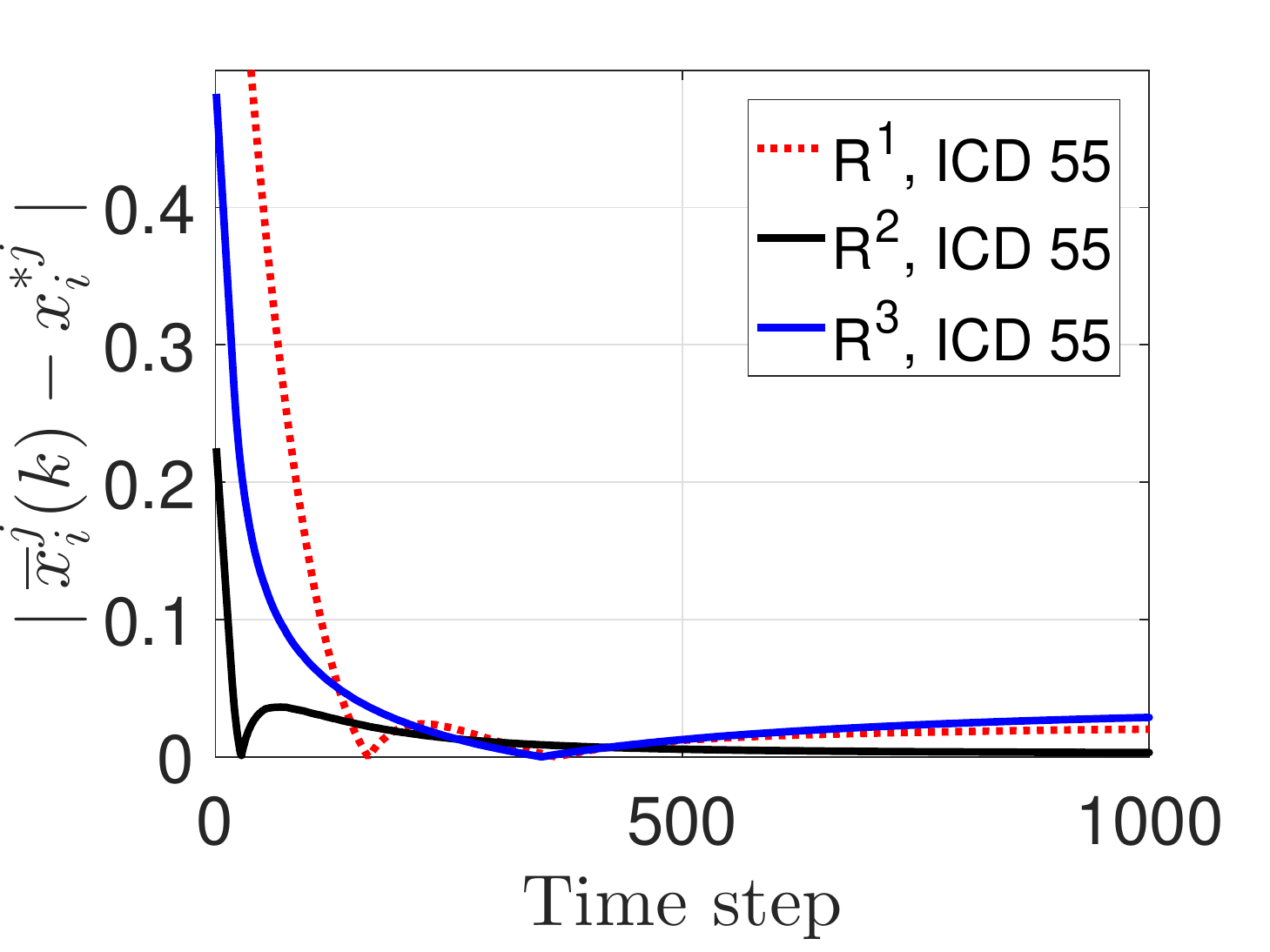}}
		\label{abs_diff_opt}\hfill
		
		\caption{Results of deterministic AIMD: (a) evolution of profile of derivatives of cost functions $\nabla_j f_i(.)$ of all ICDs of a single simulation, (b) evolution of average allocation $\overline{x}_i^j(k)$ of resources, (c) evolution of absolute difference of average allocation and optimal allocation.}
		\label{fig1} 
	\end{figure*}
	
\begin{figure*}[h] 
		\centering
		\subfloat[]{%
		\includegraphics[width=0.33\linewidth]{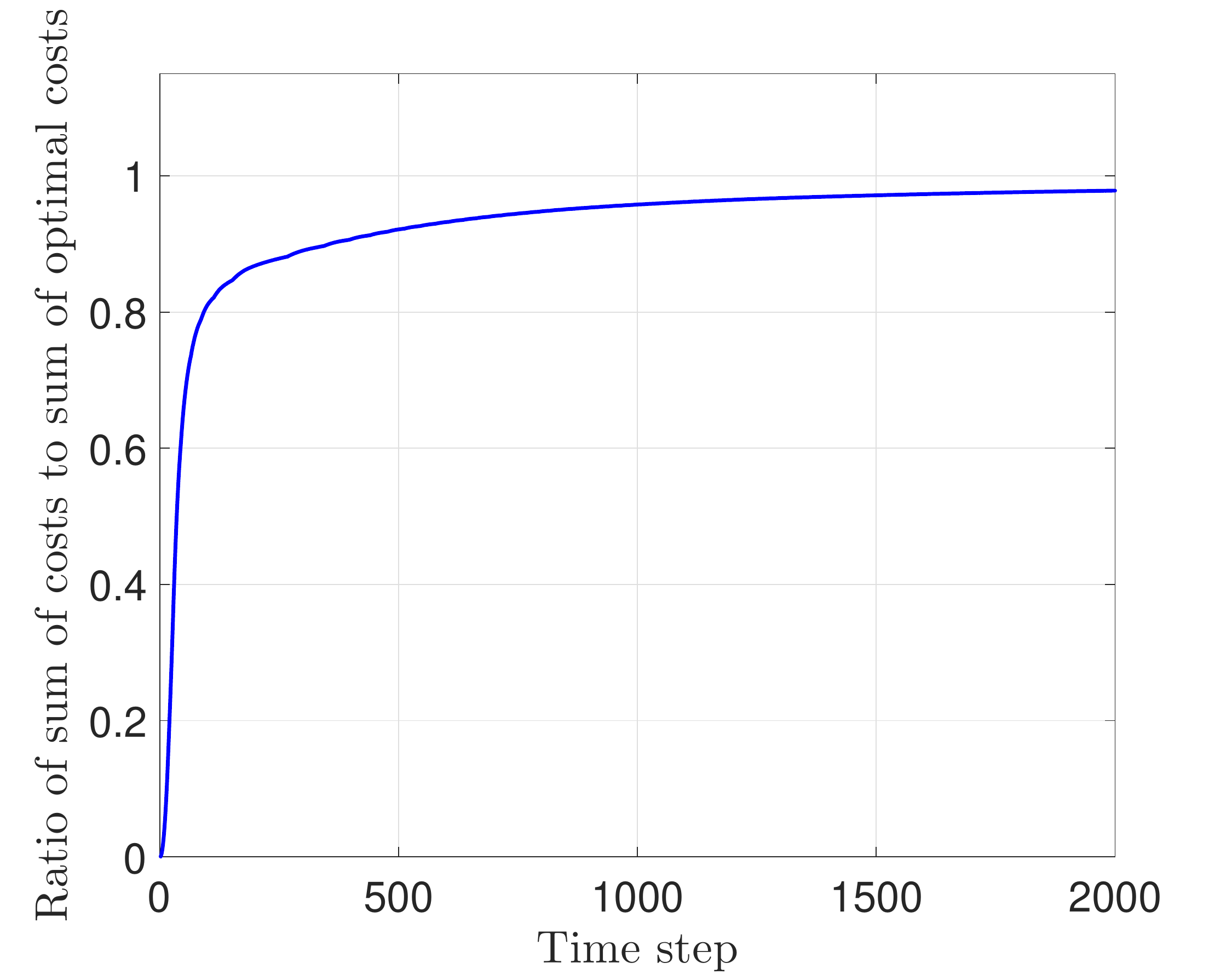}}
		\label{ratio_sum_cost_fn}\hfill
		\subfloat[]{%
			\includegraphics[width=0.33\linewidth]{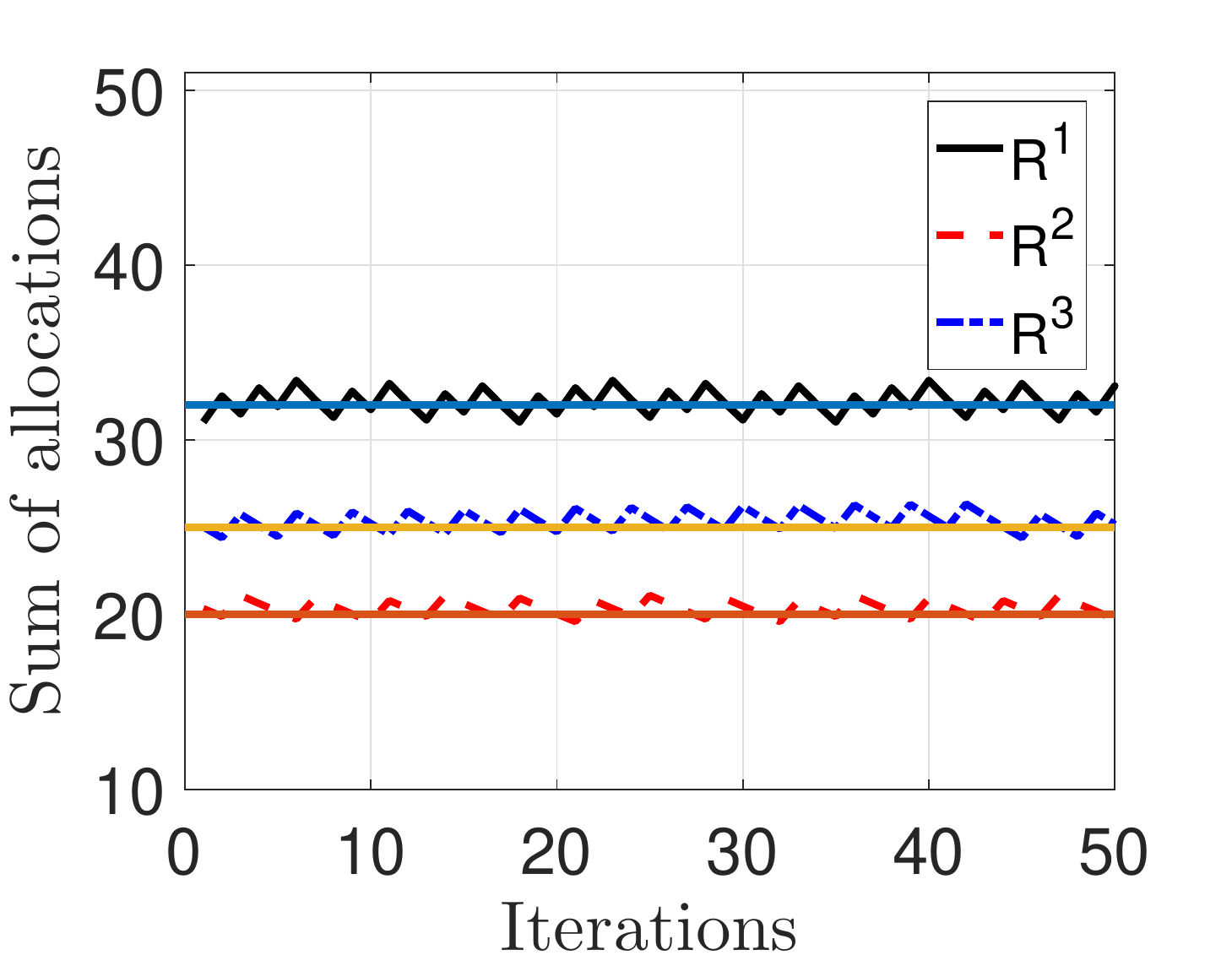}}
		\label{sum_alloc_rand_alpha2}\hfill
		\subfloat[]{%
			\includegraphics[width=0.33\linewidth]{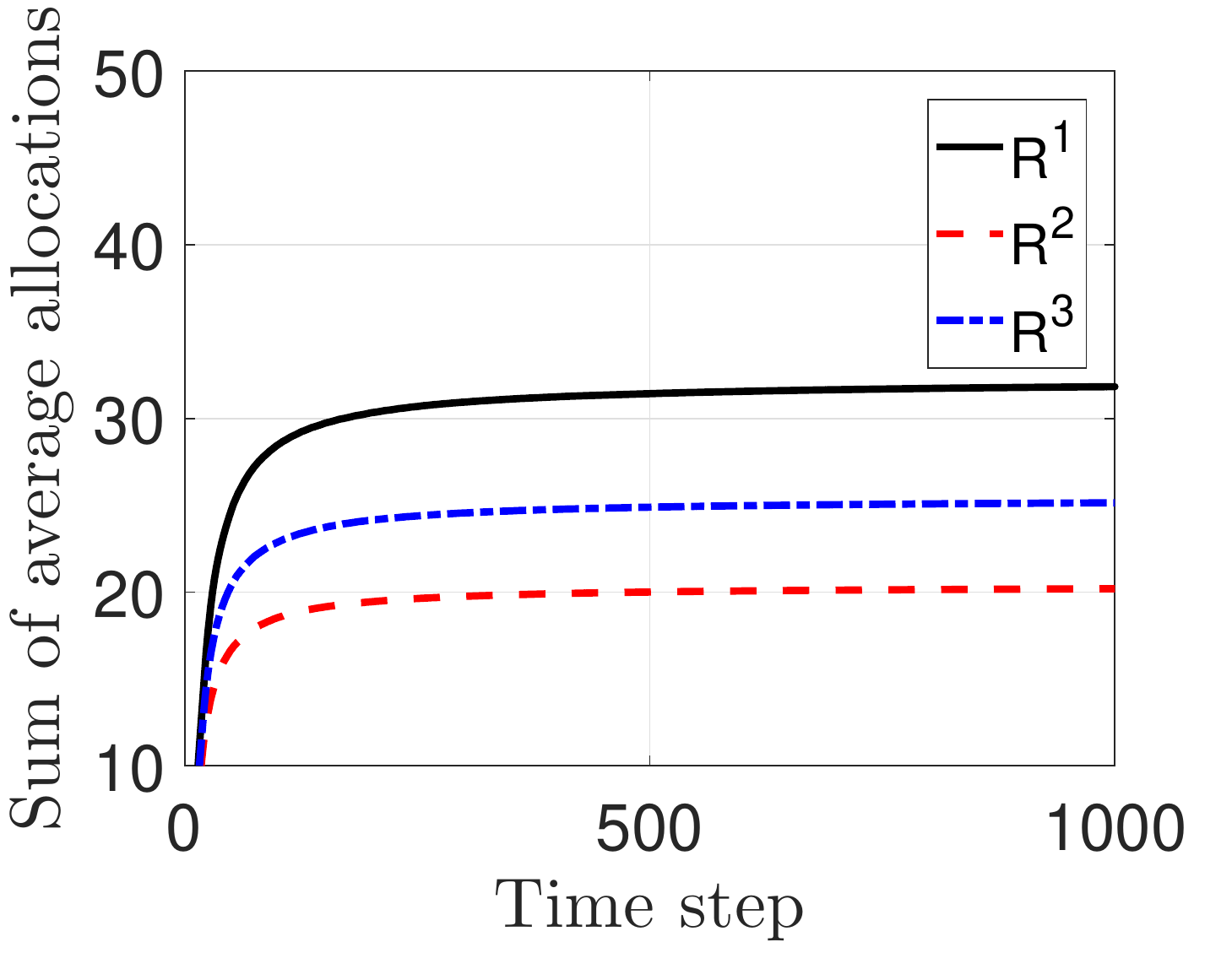}}
		\label{sum_avg_rand_alpha2}\hfill
		
		\caption{Results of deterministic AIMD: (a) ratio of the sum of cost functions to the sum of optimal cost functions, (b) total allocation of resources for last $50$ time steps, (c) evolution of sum of average allocation of resources, the capacities are $C^1=32$ GB, $C^2=20$ GHz and $C^3=25$ $\mathrm{GB^D}$.}
		\label{fig2} 
	\end{figure*}

	The following are some of the results obtained from the simulation. As described in Section \ref{prob_form}, for the optimization Problem \ref{obj_fn1}, if derivatives of cost functions with respect to a particular resource are in consensus then there exists a unique optimal solution. We observe from Figure \ref{fig1}(a) that the derivatives of cost functions of ICDs with respect to a particular resource gathers around the same value over time and hence make a consensus. Figure \ref{fig1}(b) shows that the average allocations of resources converge over time. Because the derivatives of cost functions make consensus, therefore we can say that the long term average allocations of resources are optimal allocations. We verify this claim by simulating the same optimization problem with same parameters in a centralized setting. We see that the absolute difference of the average allocation and the optimal allocation calculated in a centralized setting are close to zero (see Figure \ref{fig1}(c)). Furthermore, Figure \ref{fig2}(a) illustrates that the ratio of the sum of cost functions for average allocations and the sum of cost functions for calculated optimal allocations are approximately $1$.  
	Figure \ref{fig2}(b) shows the overshoots of the allocations, we observe that the sum of instantaneous allocations are concentrated near the respective capacities of the resources. To reduce the overshoots, we introduce a constant $\gamma^j \in (0,1)$ and update the capacity event $S^j (k+1) =  1$, when $\sum_{i=1}^{n} x_i^j(k) > \gamma^j C^j$ as described in \cite{Eqbal2017}. From, Figure \ref{fig2}(c), we observe that the sum of average allocations are approximately equal to the respective capacities of the resources, hence we say that the sum of average allocations satisfy the constraints of the optimization problem.   
		\begin{figure*}[h] 
		\centering
		\subfloat[]{%
			\includegraphics[width=0.33\linewidth]{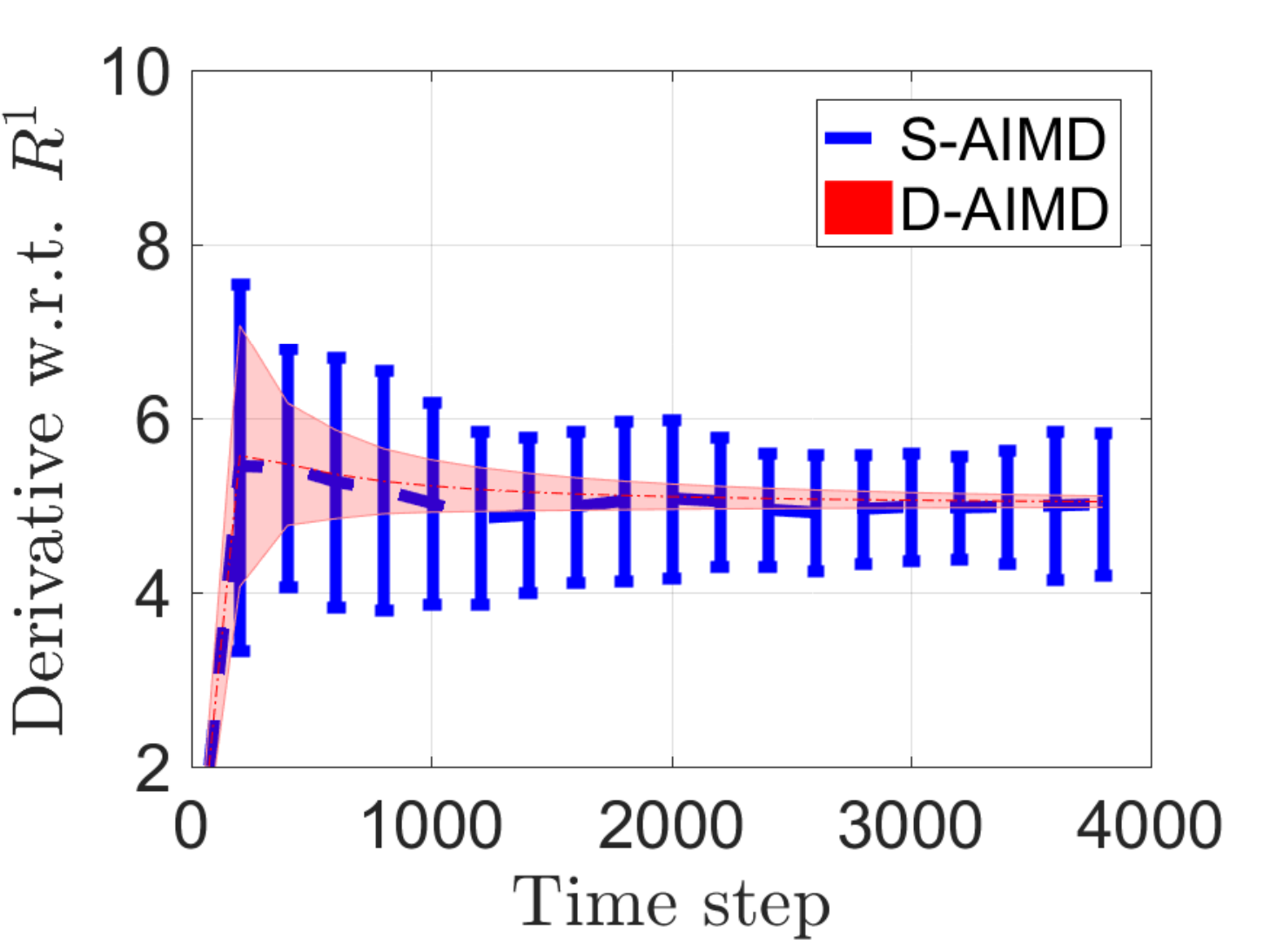}}
		\label{err_grad_R1}\hfill
		\subfloat[]{%
			\includegraphics[width=0.33\linewidth]{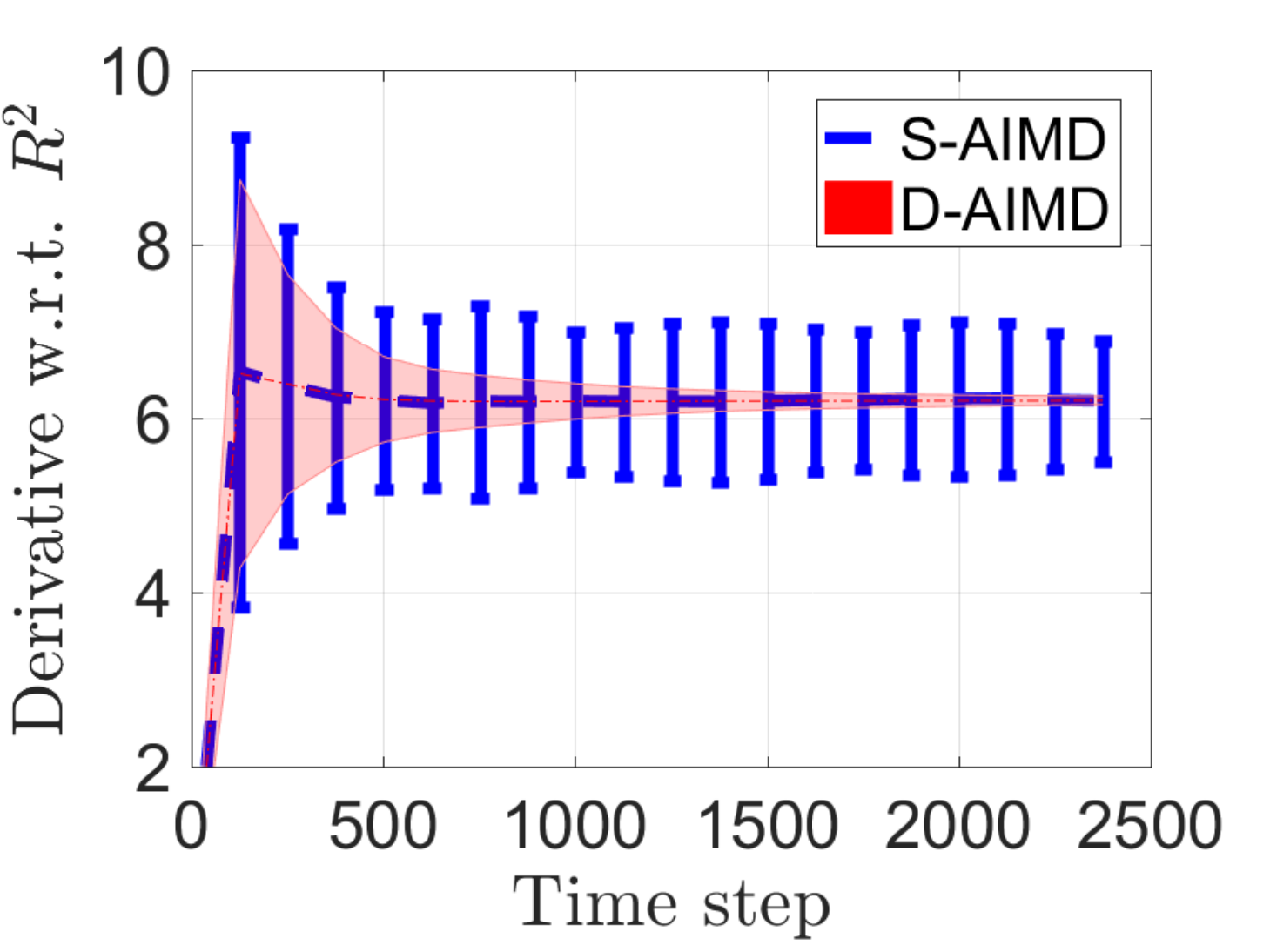}}
		\label{err_grad_R2}\hfill
		\subfloat[]{%
			\includegraphics[width=0.33\linewidth]{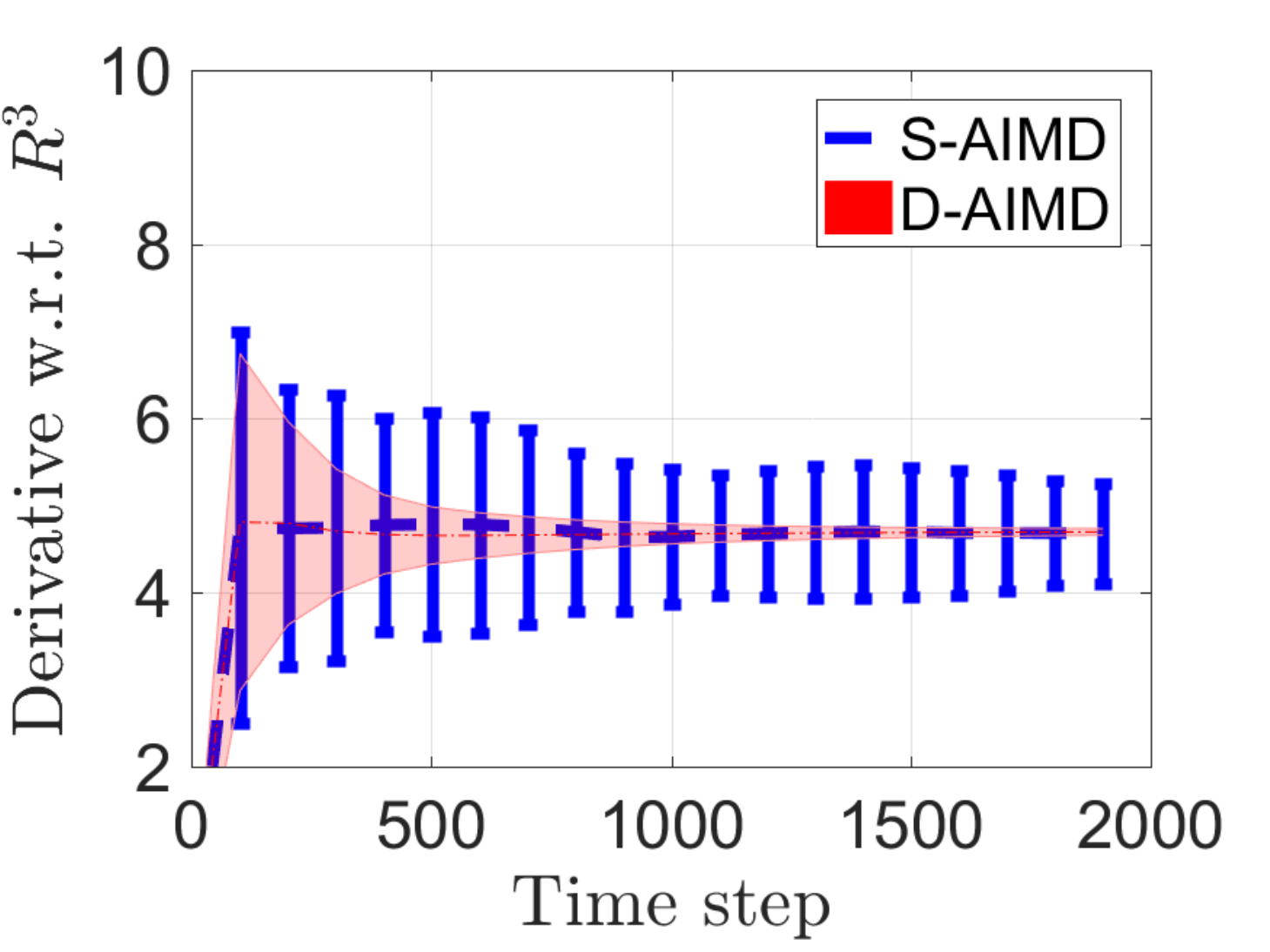}}
		\label{err_grad_R3}\hfill
		\caption{ Evolution of profile of derivatives of cost functions $\nabla_j f_i(.)$ of all ICDs of a single simulation of D-AIMD and S-AIMD algorithms with respect to --- (a)  resource $R^1$, (b) resource $R^2$, (c) resource $R^3$.}
		\label{fig3}
	\end{figure*}

\begin{figure*}[h] 
		\centering
		\subfloat[]{%
			\includegraphics[width=0.33\linewidth]{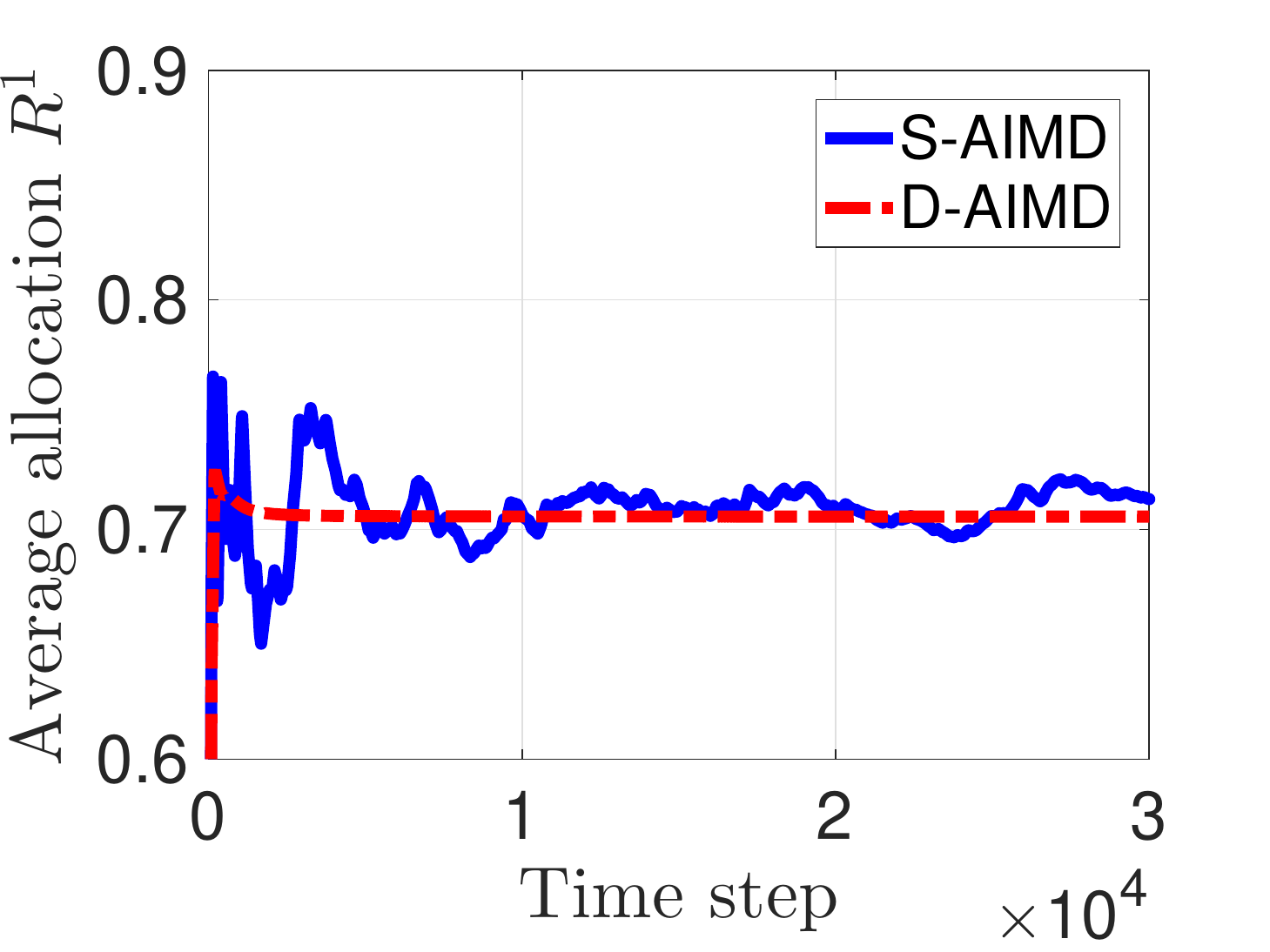}}
		\label{avg_alloc_R1}\hfill
		\subfloat[]{%
			\includegraphics[width=0.33\linewidth]{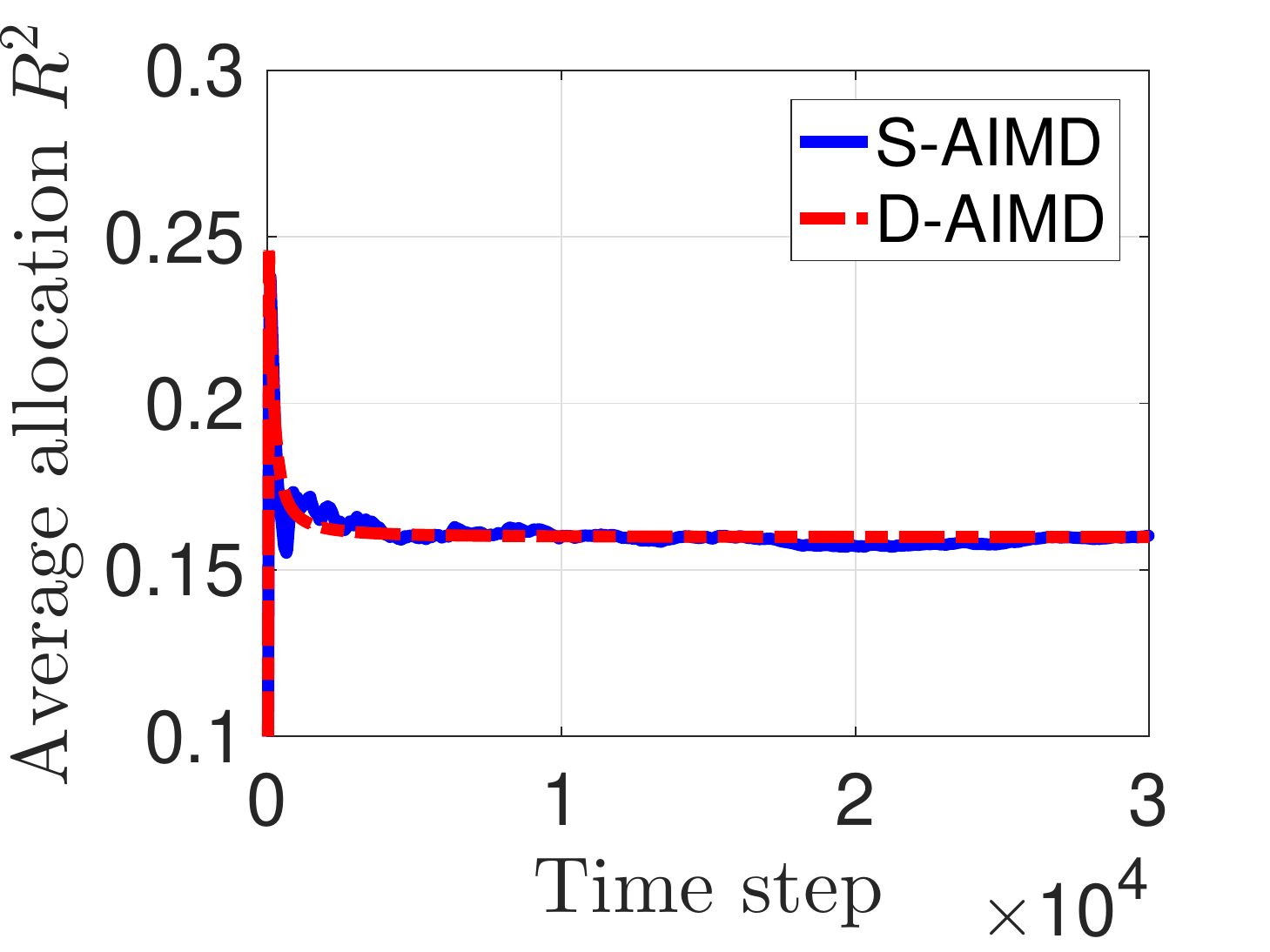}}
		\label{avg_alloc_R2}\hfill
		\subfloat[]{%
			\includegraphics[width=0.33\linewidth]{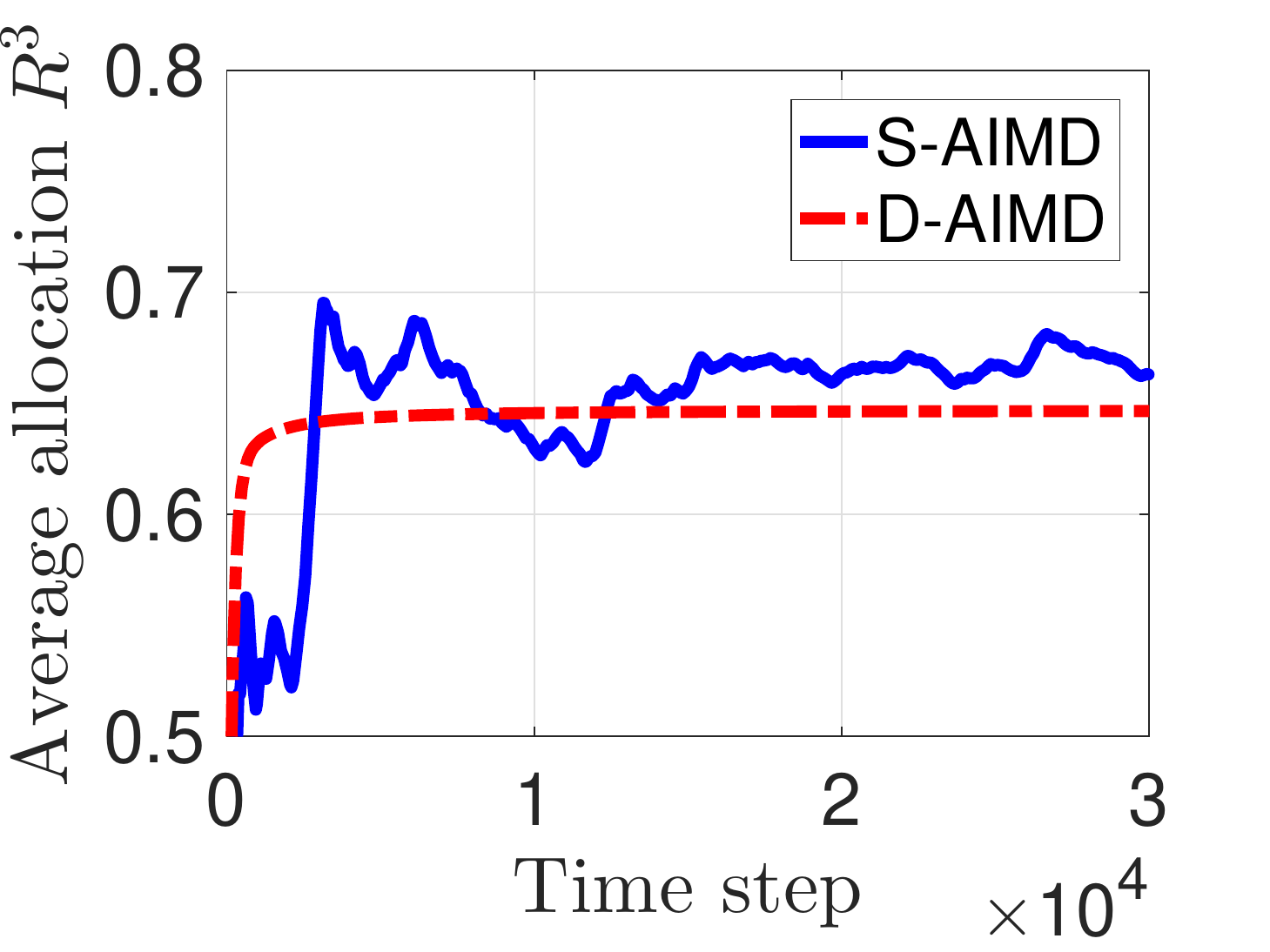}}
		\label{avg_alloc_R3}\hfill
		
		\caption{ Evolution of average allocation of resources of ICD $42$ of D-AIMD and S-AIMD algorithms --- (a)  resource $R^1$, (b) resource $R^2$, (c) resource $R^3$.}
		\label{fig6}
	\end{figure*}

	Now, we simulate stochastic AIMD approach \cite{Eqbal2017} with same parameters and cost functions. We  compare the results of deterministic AIMD approach and the stochastic approach. Figure \ref{fig3} illustrates the derivatives of cost functions with respect to resources $R^1$, $R^2$ and $R^3$, respectively of a single simulation of all ICDs of deterministic AIMD (D-AIMD) and stochastic AIMD (S-AIMD). We present the derivatives of D-AIMD as shaded error bars. We see that the derivatives of all ICDs with respect to a particular resource of deterministic AIMD algorithm concentrate more and more around the same value much faster than the derivatives obtained from stochastic AIMD algorithm. Hence, the derivatives of deterministic AIMD converge much faster than that of stochastic AIMD, therefore they reach consensus faster than stochastic AIMD. Furthermore, we observe in Figure \ref{fig6} that the average allocations of resources $R^1$, $R^2$ and $R^3$, respectively in deterministic AIMD algorithm converge faster than the average allocations in stochastic AIMD algorithm. However, the average allocations of both the approaches reach approximately same value over time. 
		\begin{figure*}[h] 
	\centering
	\subfloat[]{%
		\includegraphics[width=0.28\linewidth]{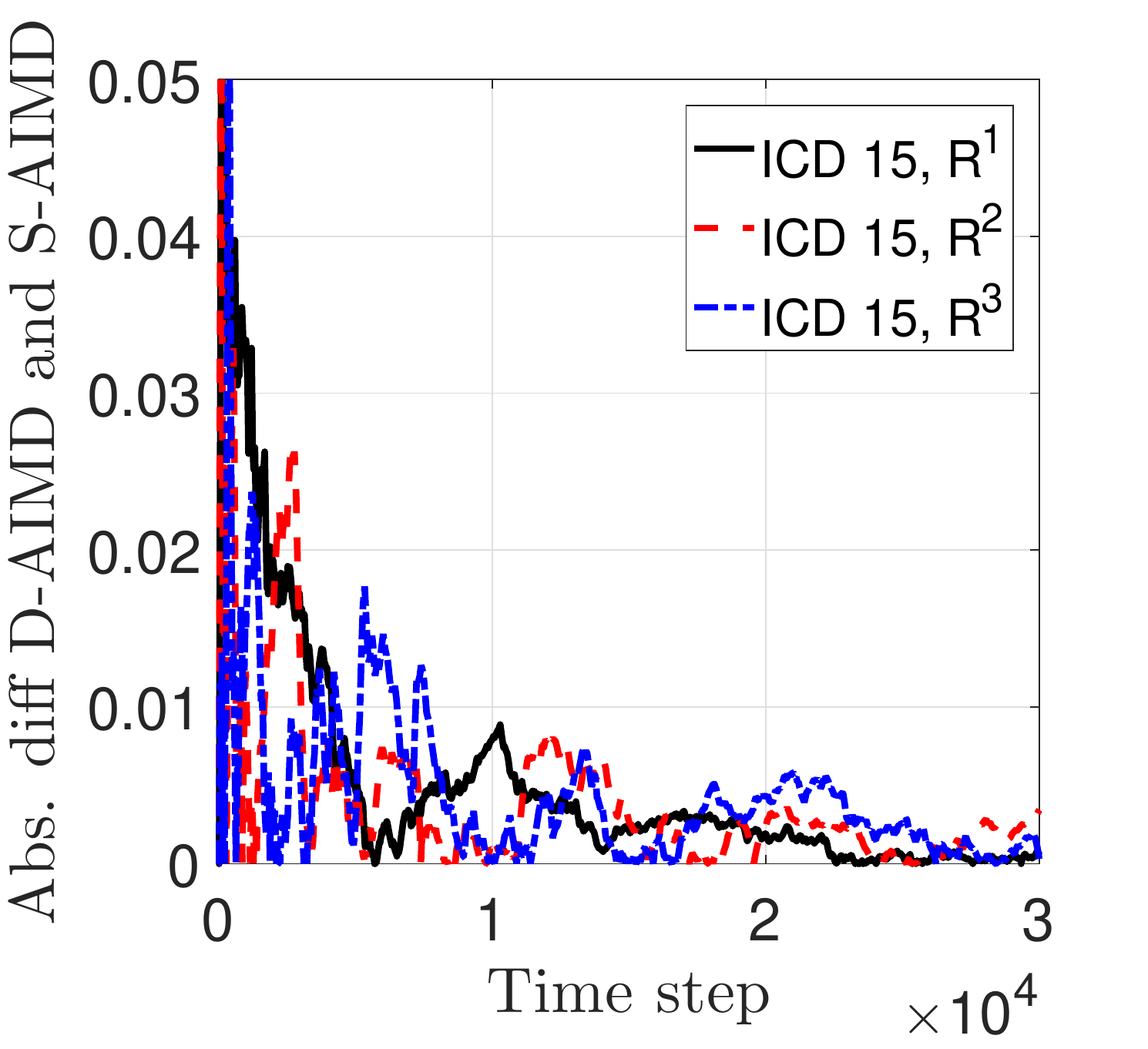}}
	\label{diff_avg_DAIMD_SAIMD}\hfill
	\subfloat[]{%
		\includegraphics[width=0.31\linewidth]{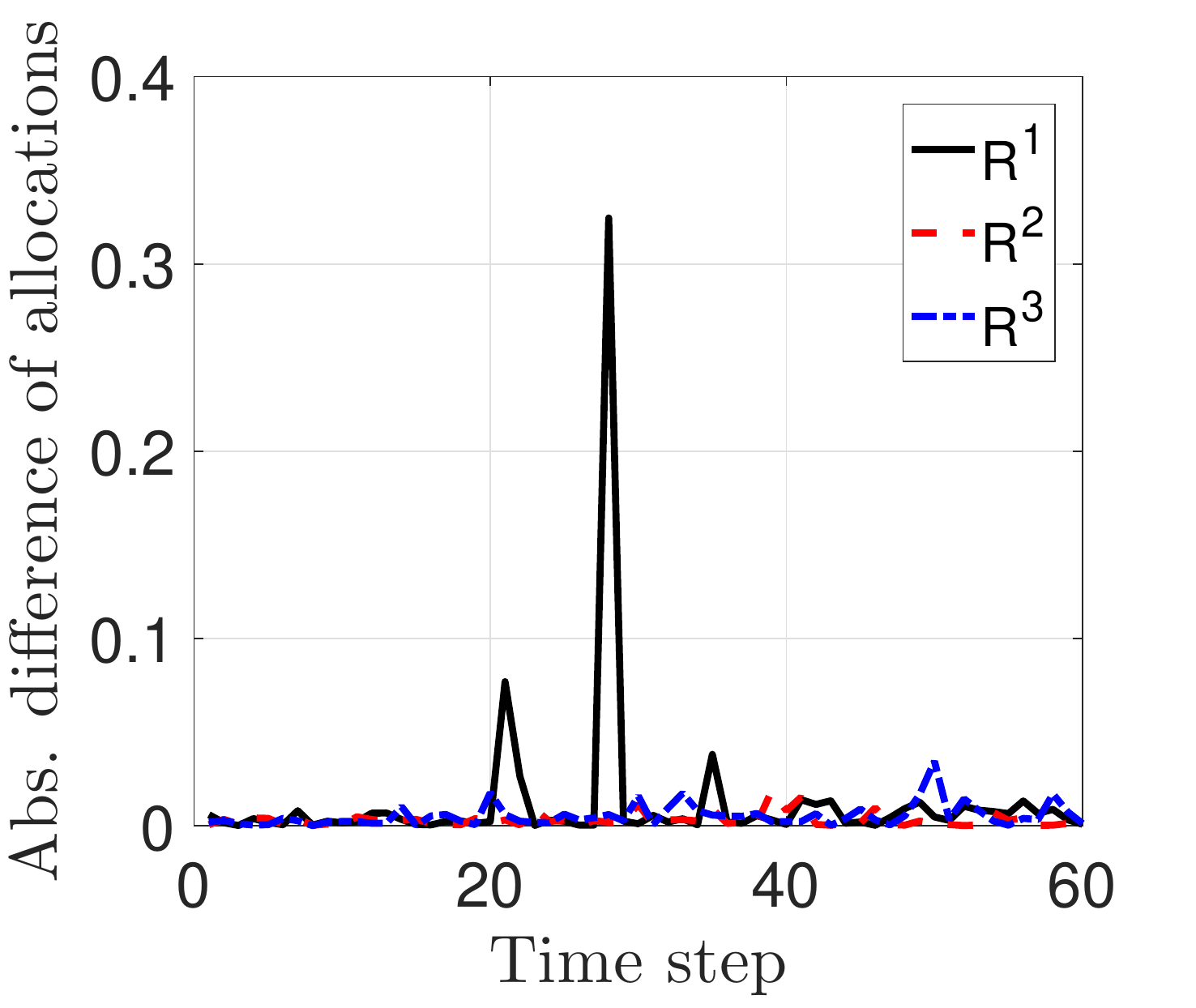}} \label{diff_avg_aloc}\hfill
	\subfloat[]{%
		\includegraphics[width=0.35\linewidth]{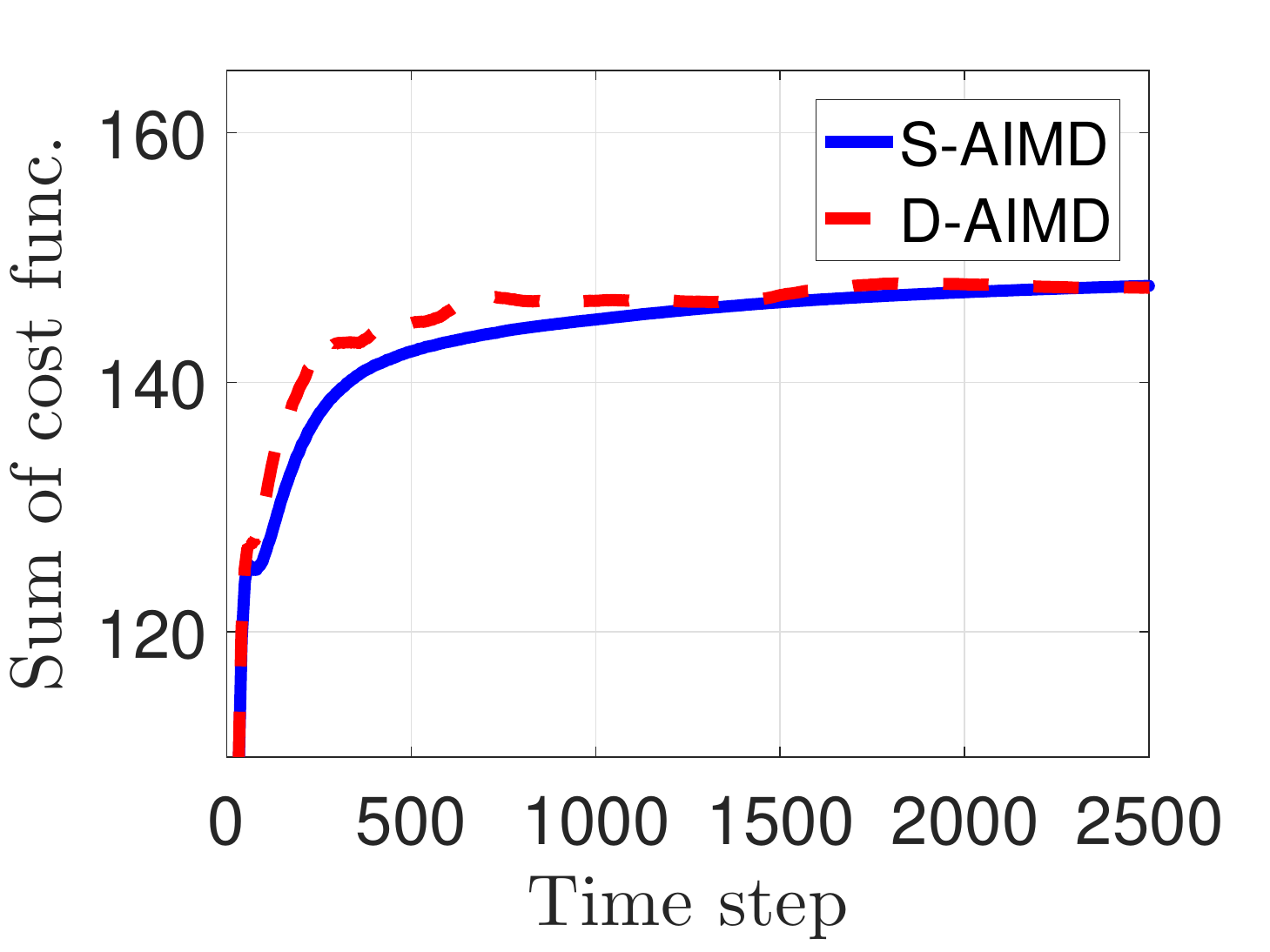}}
	\label{sum_fn_DAIMD_SAIMD}\hfill
	\caption{(a) Evolution of absolute difference of average allocations obtained from D-AIMD and S-AIMD, (b) absolute difference of average allocations obtained from D-AIMD and S-AIMD at time step $30000$, (c) evolution of sum of cost functions with average allocation of S-AIMD and D-AIMD.}
	\label{fig5}
\end{figure*}
	 This is also evident from Figure \ref{fig5}(a) which illustrates the absolute difference of average allocations obtained from D-AIMD and S-AIMD, where the absolute difference is close to zero. We observe that most of the absolute differences of average allocations reach order of $10^{-3}$ over time (Figure \ref{fig5}(b)), hence both the approaches provide approximately same long-term average allocations. Figure \ref{fig5}(c) illustrates the convergence of the sum of cost functions $\sum_{i=1}^{n} f_{i}(x_i^1(k), x_i^2(k), x_i^3(k))$ obtained from both the approaches, we observe that the sum of cost functions obtained from deterministic AIMD converges faster than sum of cost functions obtained from stochastic AIMD but they converge to same value over time. Therefore, both the approaches provide same social optimum value.
	\begin{table*}[h]
		\caption{The execution time of each simulation and the number of capacity events broadcast by control unit, the simulation is run on Intel Core i5-6500, CPU 3.2 GHz, 8 GB RAM.}
		\centering
		\begin{tabular}{|c|c|c|c|c|}
			\hline
			\multirow{2}{*}{No. of iterations} & \multicolumn{2}{c|}{No. of capacity events (bits) for $R^1,R^2,R^3$ } & \multicolumn{2}{c|}{Execution time (seconds)} \\ \cline{2-5} 
			& S-AIMD                 & D-AIMD                & S-AIMD             & D-AIMD            \\ \hline \hline
			500     & 184, 187, 137   &  266, 350, 318      & 0.666405  &   0.344306        \\ \hline
			1000    & 381, 388,  283  &  498, 719, 631      & 1.437232  &   0.681150      \\ \hline
			3000    & 1185, 1199,  835   & 1569,  2151, 1908   & 6.655815  &   3.984121         \\ \hline
			5000    & 1918,  1974,  1413  & 2439, 3563, 3288    & 16.052666  &  10.643811         \\ \hline
			10000   &  3889, 3950, 2800  & 5446, 6880,  6151   & 65.529710  &  50.616229      \\ \hline
			30000   & 11515, 11851, 8357  &  -  & 989.557591  & -       \\ \hline
		\end{tabular} \label{tab_1}
	\end{table*}
	Additionally, the execution time and the number of capacity events broadcast by control unit for different simulations is presented in Table \ref{tab_1} of both D-AIMD and S-AIMD algorithms. We would like to clarify that these results are based on simulation performed on a single computer and we do not take into account the communication delay between ICDs and control unit. We observe that the execution time of deterministic AIMD algorithm is less than that of stochastic AIMD algorithm, whereas the number of capacity events are more than stochastic AIMD for a particular simulation. Notice that, because the D-AIMD converges faster than S-AIMD, therefore in D-AIMD the control unit broadcasts overall less number of capacity events to reach consensus. Furthermore, we would like to clarify that in Table \ref{tab_1} we have not recorded the number of capacity events and execution time for D-AIMD at $30000$ time steps, because in the simulation we observe that the derivatives converge in less than $5000$ time steps.
	\section{Conclusion} \label{conc}
	We propose a distributed, private and deterministic algorithm to solve a set of distributed optimization problems for multi-resource allocation with little communication overhead. The proposed algorithm is a derandomized version of a stochastic AIMD algorithm. The solution has several features, such as the Internet-connected devices need not communicate with each other to reach social optimum. Second, the solution is communication-efficient, in the sense that the control unit broadcasts a one bit feedback signal for each resource when the demand of a resource exceeds the capacity of the resource. The solution is independent of the dimension of the system. Furthermore, failure of an Internet-connected device does not affect the performance of the system. We showed empirically that the deterministic AIMD algorithm converges faster than the stochastic AIMD algorithm. We further showed that both approaches provide approximately same long-term average allocation and achieve same social optimum cost. It is interesting to deploy the proposed algorithm in real applications using ICDs and a Cloudlet and analyze the performance of the system. It is also interesting to prove the convergence of average allocations theoretically. Additionally, the algorithm can be extended in several application areas, like smart grid, intelligent transportation systems, supply chain management, etc.
		
	\bibliographystyle{IEEEtran}
	\bibliography{DistOpt_bib}
	
\end{document}